\documentclass[11pt,a4paper]{article}
\usepackage{geometry,fancyhdr,setspace,color,parskip,eurosym,sidecap,float}
\usepackage{enumitem}
\usepackage{authblk}
\usepackage{multirow,graphicx,pdflscape,tabularx,epstopdf,siunitx}
\usepackage[position=top]{subfig}
\usepackage[labelsep=period,justification=justified,singlelinecheck=false]{caption}
\usepackage{amsmath}
\usepackage[apaciteclassic]{apacite}
\graphicspath{{Figures/}}

\begin{document}

\title{Beyond synchronization: Body gestures and gaze direction in duo performance}
\author[1,2]{Laura Bishop}
\author[1]{Carlos Cancino-Chac\'on}
\author[3]{Werner Goebl}
\affil[1]{Austrian Research Institute for Artificial Intelligence (OFAI), Vienna, Austria}
\affil[2]{RITMO Centre for Interdisciplinary Studies in Rhythm, Time and Motion, University of Oslo, Norway}
\affil[3]{Dept. of Music Acoustics--Wiener Klanstil, University of Music and Performing Arts Vienna, Austria}
\date{}
\maketitle

\vspace{-9cm}
\scriptsize Bishop, L., Cancino-Chac\'on, C., \& Goebl, W. (2021). Beyond synchronization: Body gestures and gaze direction in duo performance. In Timmers, R., Bailes, F., and Daffern, H. (Eds.), Together in Music: Participation, Co-Ordination, and Creativity in Ensembles. Oxford: Oxford University. This version is a preprint of the chapter prepared by the authors. \normalsize
\vspace{7.5cm}

Successful coordination during music ensemble performance is supported by dynamic, multimodal interactions among ensemble members. Most critical is the musicians’ attention to each other’s audio signals: this alone is sufficient for an ensemble to maintain synchronization much of the time (see chapter 17). Visual interaction is largely peripheral to successful synchronization, though performers may choose to exchange visual signals when timing of the music is uncertain.

While successful synchronization provides a foundation for a high-quality performance, coordination of other parameters is important as well. In the Western classical music tradition, expressive parameters such as dynamics and phrasing must be carefully aligned. Coordinated manipulation of such parameters – which are imprecisely specified (or omitted) in the score, is important for ensembles who want to establish an interpretation of the piece that distinguishes their unique group playing style.

What role visual interaction plays in coordinating an expressive performance remains unclear. To date, few studies have considered the possibility that its contribution may be social, rather than related to the basic mechanics of synchronization. We propose that visual interactions among ensemble performers may help with achieving complex expressive goals, and more specifically, may serve as a social motivator that allows performers to confirm and guide each other’s attention. Ultimately, this may promote creative risk-taking. 

In this chapter, we focus on two main categories of visual interaction: body gestures and gaze direction. Our focus on body gestures is motivated by research showing that gesture patterns often change during joint action tasks to become more predictable (van der Wel et al., 2016; see also chapter 14). Moreover, coordination sometimes emerges between musicians at the level of body sway (Chang et al., 2017). Our focus on gaze direction was motivated by the fact that gaze can serve simultaneously as a means of obtaining information about the world and as a means of communicating one’s own attention and intent.

We carried out a study with musical duos to test two broad, competing hypotheses. On one hand, we considered the possibility that visual interaction supports basic coordination as a supplement to auditory interactions, resulting in musicians being more visually interactive during difficult-to-coordinate parts of a performance and during the early stages of rehearsal, when they are unsure of how each other might play. The competing hypothesis is that visual interaction supports engagement and creativity among co-performers, resulting in musicians being more visually interactive during the later stages of rehearsal, when performers are more sure of the notes and more confident about experimenting with new interpretive ideas. Our methods and results are presented in detail elsewhere; see Bishop et al. (2019a, 2019b). The current chapter draws our results together into a discussion of how visual interaction may enhance performers’ experience of playing together and audiences’ perceptions of performance quality.    

\begin{figure}[h!]
\centering
{\includegraphics[width=.9\textwidth]{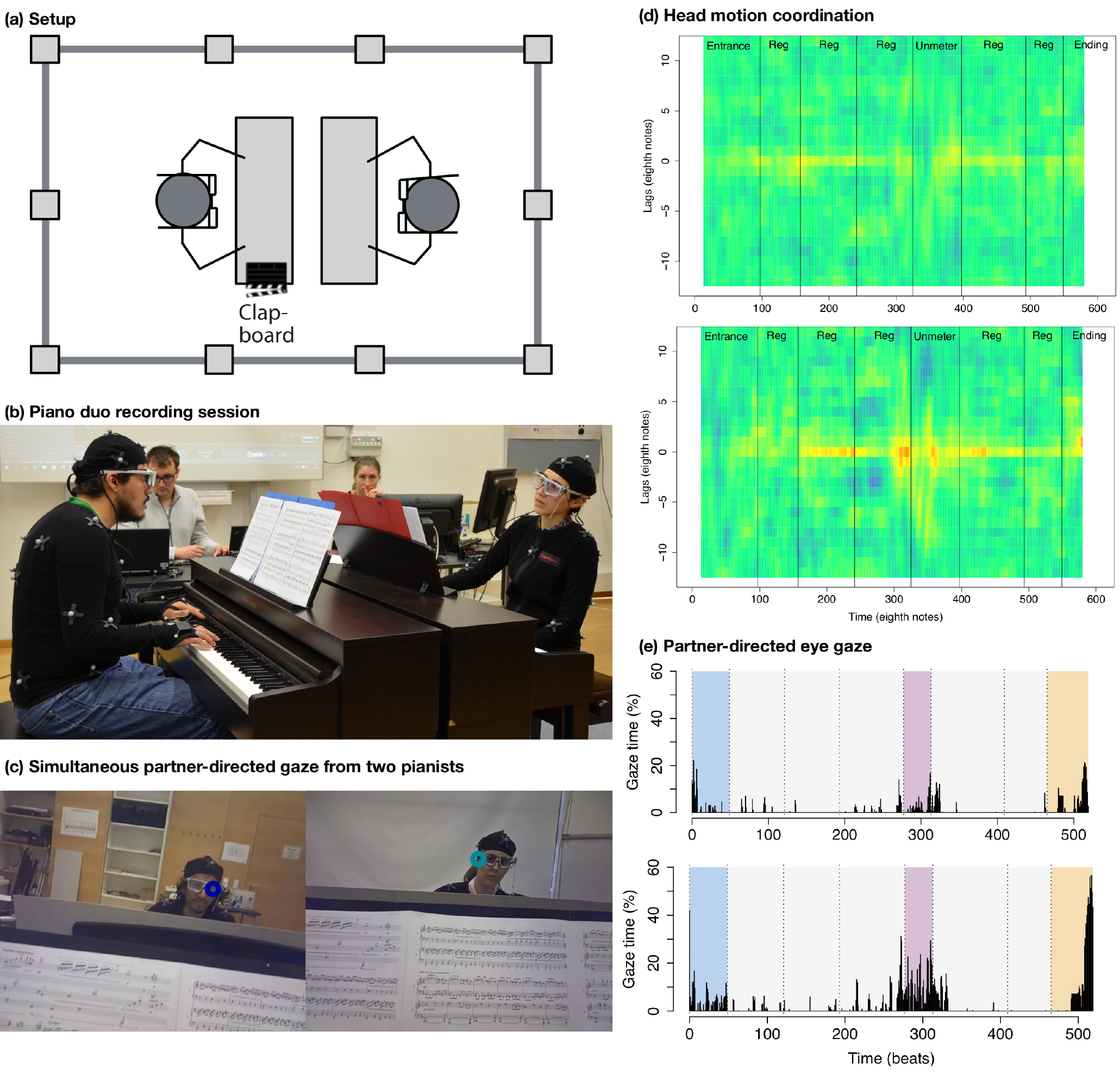}}
\caption{Illustration of research methods. (a) The experimental setup. (b) Pianists are shown performing with motion capture and eye tracking. (c) Video captured by the eye tracking glasses is shown for two pianists; blue circles indicate gaze position. (d) The strength of primo-secondo head coordination, measured using a rolling-window cross-correlation method and averaged across pianists, for the 1st (upper plot) and 3rd (lower plot) performances; blue areas indicate negative correlations and red areas indicate positive correlations. (e) The percentage of performance time that pianists spent watching their partner in the 1st (upper plot) and 3rd (lower plot) performances; grey dotted lines indicate metrical boundaries, the beginning and ending piece sections are in blue and yellow, and the unmetered section is in pink. Note that beats were counted differently for plots in (d) and (e), resulting in different total beat numbers. (Images were published previously in Bishop et al., 2019a, Bishop et al., 2019b).}
\end{figure}

\section*{Methods and Results}

Ten clarinet duos and ten piano duos, all comprising highly-trained professional or semi-professional musicians, participated in the study. Each duo spent one hour rehearsing a new piece, and recorded four complete performances of the piece during the course of the rehearsal period. 

At the start of the session, musicians were instructed to practice together rather than independently, and to focus on preparing an expressive interpretation of the piece. We pointed out that some elements of the score were deliberately ambiguous (e.g., there were no tempo ranges given), and told the musicians to make their own decisions about how the piece should sound. They were told that the purpose of the study was to look at how musicians interact during rehearsal, and were not given any specific instructions regarding where to look or how to move.

The piece was specially composed for the purpose of the study, and contained a number of potential challenges for coordination, including a section with no notated meter. Motion capture recordings were collected using a 10-camera OptiTrack system, and SMI wireless eye tracking glasses were used to collect gaze data. Audio and (for pianists) MIDI data were also recorded. 

Duos were arranged face-to-face, a short distance apart, and were allowed to use the score throughout the session (Figure 1a—c). Duos performed the piece once at the start of the session (i.e., while sight-reading), once halfway through a free joint rehearsal period, and twice at the end of the session. The fourth performance was given under no-visual-contact conditions (i.e., performers were blocked from seeing each other); all other performances were given under normal visual contact conditions.

\subsection*{Effects of musical structure}

We investigated the effects of musical structure on eye gaze patterns and head kinematics. Of primary interest was the comparison between the unmetered section of the piece, where timing was fairly irregular, and the metrical, more regularly-timed sections. The unmetered section allowed for greater freedom of interpretation, and was likely more difficult for duos to coordinate.

Coordination of head motion between performers was stronger during the unmetered section than elsewhere in the piece. In Figure 1d, this is shown by the uptake in red and yellow colouring that occurs in the unmetered section around lag 0, and is particularly visible in the 3rd performance (lower plot). Also notable in the 3rd performance is the increase in coordination strength (i.e., red uptake) that occurs immediately prior to the unmetered section, which likely indicates a coordinated transition between sections or preparation for the less-predictable unmetered section.

Performers were also found to exchange cueing gestures (exaggerated head nods) during the unmetered section. Drawing on some previous studies of ours, which showed that beats are primarily communicated through gesture acceleration (Bishop \& Goebl, 2018), we conducted a search for cueing gestures by comparing head acceleration curves at each beat of the piece to head acceleration curves given just prior to piece onset (at the moment where cueing-in gestures were exchanged). Acceleration curves more closely replicated those associated with cueing-in gestures at fermatas, suggesting that performers exchanged visual signals to help align the chords that followed these pauses.

Musicians spent a larger percentage of performance time watching their partner during the unmetered section than during the regularly-timed sections, though most musicians did occasionally look at their partner during regularly-timed sections as well (Figure 1e). There was also an uptake in partner-directed gaze at the very start of the piece (which started with a synchronized chord) and in the final beats of the piece (where most duos introduced a ritardando). Also, importantly, we found no evidence that the “leader” (i.e., the partner playing the melody) looked less at the “follower” (i.e., the partner playing the accompaniment) than the follower spent looking at the leader. This suggests that visual information travelled both ways, rather than exclusively from leader to follower.

\subsection*{Effects of rehearsal}

The effects of rehearsal on eye gaze and head kinematics were also assessed. Our analyses of head gestures revealed that quantity of movement and the strength of coordination between performers were higher at the end of the session than at the start (Figure 1d). The percentage of time that musicians spent watching their partner also increased from the start to the end of the session (Figure 1e).

\subsection*{Effects of visual contact}

A comparison between the final two performances was made for gesture kinematics, to test whether head movements were more communicative when musicians could see each other than when visual contact between performers was occluded. We found that quantity of movement and the strength of coordination between performers' head movements were both greater when musicians could see each other than when they could not. 

The quality of synchronization that musicians achieved during their final two performances was also assessed, by looking at the magnitude of primo-secondo note onset asynchronies. Synchronization did not differ between performances, indicating that visual contact – and the strengthened gestural coordination that occurred during the third performance – had no noticeable effects on synchrony.

\section*{Discussion}

Our findings show how duo musicians' use of visual interaction techniques changes across different performance conditions. When predictability of timing is low (as during our unmetered section), performers seem motivated to supplement their auditory interactions with visual signals. Cueing gestures were exchanged as performers tried to re-synchronize following fermatas. Performers also spent more time watching each other during the unmetered section. These findings suggest that visual signals may sometimes be used to facilitate coordination. On the other hand, we observed no difference in synchronization success between the final two performances (with vs. without visual contact). Thus, we might predict that performers' use of visual signals during the unmetered section does more to raise their confidence than it does to affect coordination.

Performers moved more, and their movements were more coordinated after rehearsing than before; likewise, the time they spent watching each other increased. Performers were therefore more visually interactive when they were more familiar with the music (and their partner). Musicians' expectations for how the piece should sound may be more concrete after some rehearsal, and their visual interaction may reflect a more driven attempt to ensure successful coordination. A second (not mutually exclusive) explanation is that once musicians are more comfortable with the notes, their focus shifts more towards their partner, with whom they become more engaged, making both partners more willing to experiment with alternate interpretations. In this respect, visual interaction may come to serve a social-motivational function. 

Our hypothesis that visual interaction among ensemble members may serve a primarily social function is motivated by recent research on group flow, a state characterized by a shared sense of absorption, intrinsic reward, and effortlessness (Hart \& Di Blasi, 2015).  Musicians associate a lack of self-consciousness and individuality with the experience of group flow, but also require a clear understanding of their role in the collaboration. The possibility for dynamic, fine-grained interaction between musicians is thought to be critical for establishing a sense of “shared social presence”. We propose that visual interaction may encourage an external (partner/task-directed) rather than internal (self-reflective) focus and strengthen the resonance that emerges between co-performers. 

We also propose that visual interaction may benefit performance quality from the audience's perspective. Some research has shown that viewers are sensitive to the social dynamics portrayed by interacting musicians (e.g., dominance or insolence); however, further study is needed to determine which types of (audio and visual) cues are meaningful to audiences.

During public performance, the audience's overt response can feed back to the performer, shaping their performance as it unfolds. Performer-audience interaction is present during Western classical concerts, though perhaps less obvious than in some other musical scenes. We would argue that musicians trained in the Western classical tradition (where public performance is typically the goal) are even in rehearsal influenced by the response of a hypothetical audience. Thus, musicians may “play for” this hypothetical audience during rehearsal, as they anticipate potential audience responses. We would therefore hypothesize that our participants' recorded performances included a similar, if not underestimated (due to lowered arousal) degree of visual interaction than would occur among duo musicians during public performance.

The current study was designed to maximize ecological validity within a laboratory setting, but the question remains: how generalizable are our results to real-world performances? Musicians encounter variability in performance conditions on a regular basis (e.g., in acoustic conditions, lighting conditions, positioning and spacing on a stage, the presence/absence of an audience, etc.). Skilled musicians are expected to adapt to these variable conditions, and indeed, tend to do so successfully. This means, of course, that researchers run the risk of capturing musicians' abilities to adapt to a new but artificial set of constraints, rather than their “real-world” behaviour. We would encourage other researchers to test the replicability of our findings with musicians of different traditions, playing different instruments, and positioned relative to each other in different ways.

\section*{Conclusions}

This study tested two potentially-complimentary functions of visual interaction in a duo performance setting. Our results suggest that skilled musicians draw on visual signalling when trying to coordinate irregularly timed music. More generally, visual interaction seems to serve a social-motivational function, as musicians seem eager to engage visually with each other, despite this offering no benefit to synchronization. Future research should consider how visual interaction affects performance quality beyond the level of synchronization, and in particular, how it shapes the subjective experiences of performers and audience members.

\section*{Acknowledgements}

This research was funded by the Austrian Science Fund (FWF), grant P-29427, and the European Research Council (ERC), under the EU's Horizon 2020 Framework Programme, grant 670035 (project “Con Espressione”).

\section*{References}

Bishop, L., Cancino-Chacón, C. E., \& Goebl, W.  (2019a).  Eye gaze as a means of giving and seeking information during musical interaction. Consciousness and Cognition, 68, 73-96. doi: 10.1016/j.concog.2019.01.002

Bishop, L., Cancino-Chacón, C. E., \& Goebl, W. (2019b). Moving to communicate, moving to interact: Patterns of body motion in musical duo performance. Music Perception, 37, 1-25.  doi: 10.1525/mp.2019.37.1.1

Bishop, L., \& Goebl, W. (2018). Communication for coordination: Gesture kinematics and conventionality affect synchronization success in piano duos. Psychological Research, 82, 1177-1194.  doi: 10.1007/s00426-017-0893-3

Chang,  A., Livingstone, S. R., Bosnyak, D. J., \& Trainor, L. J. (2017). Body sway reflects leadership in joint music performance. Proceedings of the National Academy of Sciences, 114(21). doi: 10.1073/pnas.1617657114

Hart, E., \& Di Blasi, Z. (2015). Combined flow in musical jam sessions: A pilot qualitative study. Psychology of Music, 43(2), 275-290.  doi: 10.1177/0305735613502374

van der Wel, R. P. R. D., Sebanz, N., \& Knoblich, G. (2016). A joint action perspective on embodiment.  In Y. Coello \& M. Fischer (Eds.), Foundations of Embodied Cognition. Psychology Press.

\end{document}